\documentclass[10pt,twocolumn]{article}
\usepackage{amsmath,subfig,graphicx}
\usepackage{authblk}
\usepackage[margin=2.5cm]{geometry}
\begin{document}
    \title{Quantitative flow visualization by hidden grid background oriented schlieren}

    \author[1]{Jagadesh Ramaiah}
    \affil[1]{Department of Electrical Engineering, Indian Institute of Technology Kanpur, Kanpur-208016, India }
    \author[2]{Tullio de Rubeis}
    \affil[2]{DIIIE, University of L’ Aquila, P. le E. Pontieri 1, L’ Aquila, 67100, Italy}
    \author[1,3,5]{Rajshekhar Gannavarpu}
    \affil[3]{Center for Lasers and Photonics, Indian Institute of Technology Kanpur, Kanpur-208016, India}
    \author[2,4]{Dario Ambrosini}
    \affil[4]{ISASI-CNR, Institute of Applied Science and Intelligent Systems, via Campi Flegrei 34, Pozzuoli (NA), 80078, Italy}
\affil[5]{Email: gshekhar@iitk.ac.in}
\date{}
    \maketitle
    \begin{abstract}
        The paper introduces hidden grid background oriented schlieren for quantitative  study and visualization of natural convection heat transfer.
        In this technique, the refractive index variation, induced by the temperature gradient, is encoded in the recorded signal phase through the distortion of a background pattern.
        The background (undistorted) pattern is implicit (or hidden) in the light source.
        Quantitative estimation of the phase map is obtained by windowed Fourier transform.
        This method offers localized processing of the signal using joint space-frequency representation.
        The performance of hidden grid background oriented schlieren is practically demonstrated  by investigating natural convective flow, a demanding task due to its comparatively small heat transfer.
    \end{abstract}

\section{Introduction}
Optical measurement techniques play an essential role in heat flow analysis and visualization because of their non-invasive nature, full-field measurement capability, and high sensitivity  \cite{merzkirch2012flow,ambrosini2006optical,prenel2012flow}.
The idea of mapping refractive index fields by the distortion of a background pattern is old, simple but effective. It is well known that a refractive index gradient has two effects on a beam of light passing through it: a phase change and a ray bending. Different methods can be used to reveal and measure the two effects; interferometric techniques are able to measure directly the phase change, while methods such as schlieren \cite{settles2001schlieren} are based on beam deflection evaluation. If a pattern is seen through the refractive index field, the beam deflection can be considered as a shift (and hence a deformation) of the background. 
Robert Hooke was the first to observe the background distortion effect of transparent refractive media in 1665 \cite{hooke,rienitz1997optical}, while one of the first application of background distortion technique is 1928 Lamm's method \cite{lamm1928bestimmung,lamm1936determination}, in which a scale is imaged through a diffusion cell. During the mass transfer phenomenon, the scale is distorted because of the different deflections of the light beams travelling through the cell.
A solid background to the schlieren techniques was given by Hubert Schardin \cite{schardin1942schlierenverfahren}. Unfortunately, his original 1942 German publication is difficult to find. Some information about the different canonical arrangements he proposed, including background distortion methods, can be found in Settles \cite{settles2001schlieren}.
The idea of deducing information about a refractive index field (and, hence, about the phenomenon that caused the change in the index of refraction) reappeared from time to time \cite{kopf1972application,giglio1981white} but flourished from the end of 1990's years. Re-introduced by different researchers, it was named \textit{fringe projection} \cite{massig1999measurement,perciante2000visualization,ambrosini2000flow}, \textit{synthetic schlieren} \cite{sutherland1999visualization,dalziel2000whole} and \textit{background oriented schlieren (BOS)} \cite{raffel2000applicability,richard2001principle}, with the latter now firmly established \cite{raffel2015background, settles2017review}. The work by K{\"o}pf \cite{kopf1972application} is an early example of speckle photography \cite{fomin1998speckle}, a technique that is a close relative of BOS.
Background oriented schlieren is a popular optical technique for flow visualization because of its positive features, such as robustness towards external disturbances, digital data processing, ease of use, and cost-effective design. 
In the last 20 years an abundant literature about BOS has been produced \cite{raffel2015background, settles2017review}; background patterns can be grids and/or lines \cite{sutherland1999visualization,dalziel2000whole}, sometimes obtained by interference \cite{ambrosini2007heat}, masks of dots (regular or random) \cite{raffel2000applicability}, colored dots \cite{sourgen2012reconstruction}, natural background \cite{hargather2010natural}, speckle \cite{kopf1972application,giglio1981white,ambrosini2010white,meier2013improved,nakamura2021speckle}, and gradient patterns \cite{mier2016color}.
BOS relies on very simple equipment: basically, only a background pattern and a camera are needed; equipment reduces to camera only when using natural background \cite{hargather2010natural}.
Consequently, the role of data processing methods becomes paramount for flow visualization. Typically, data processing in BOS is performed by cross-correlation \cite{raffel2015background} but also phase extraction has been proposed \cite{ambrosini2012role}.
Phase shifting \cite{zhu2018quantitative,shoji2015high,zuo2018phase} is a prominent technique for phase extraction in the domain. However, it relies on capturing multiple phase shifted images, which limits the application for dynamic investigations.
For single frame analysis, Fourier transform \cite{takeda1982fourier,ambrosini2007heat} is a popular technique for heat flow visualization; however, the method relies on global processing of the intensity signal and is thus susceptible to noise. 
Over the years, a robust data processing method, namely the windowed Fourier transform\cite{kemao2007two,kemao2004windowed} method has found diverse applications in the domain of optical metrology including non-invasive surface profilometry \cite{ri2018comparative}, non-destructive deformation testing \cite{ajithaprasad2018non} and defect identification \cite{ajithaprasad2019defect,qian2005fault}.
The main principle of operation of the windowed Fourier transform method is the local processing of the distorted intensity pattern signal by selectively choosing the spatial window and spatial local frequencies.
This joint space-frequency representation allows for robust processing of the input signal and minimizes the global propagation of detrimental effects of noise.

In this work, we couple the robust data processing capabilities of the windowed Fourier transform with hidden grid background oriented schlieren. This new variation of BOS exploits a grid implicit in the light source as a background, therefore our equipment implies only a camera and a suitable source.
To the best of our knowledge, such an experimental demonstration of the method has not been hitherto investigated for heat transfer problems.

The paper is organized as follows.
The details about the experimental setup and data processing by the windowed Fourier transform are outlined in Section 2. 
The results are presented in Section 3.
This is followed by discussions and conclusions.

\section{Experiment setup and data processing}
\begin{figure*}[t] 
    \centering
    {\includegraphics[width=0.95\textwidth]{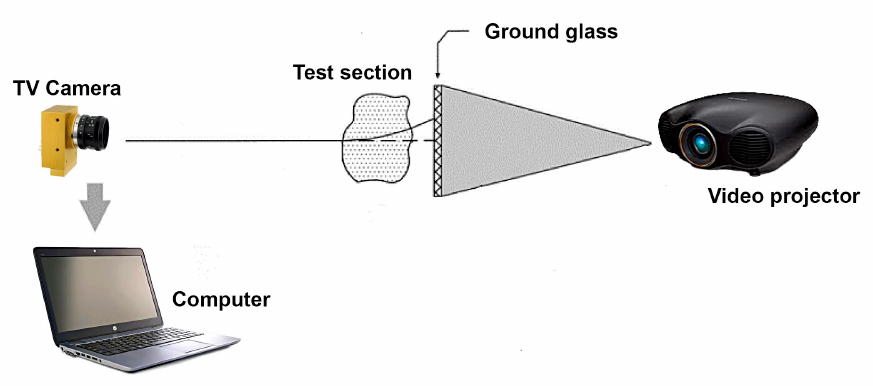}}
    \caption{Schematic drawing of hidden grid background oriented schlieren}
    \label{fig:1}
\end{figure*}

\begin{figure}[t]
    \includegraphics[width=0.45\textwidth]{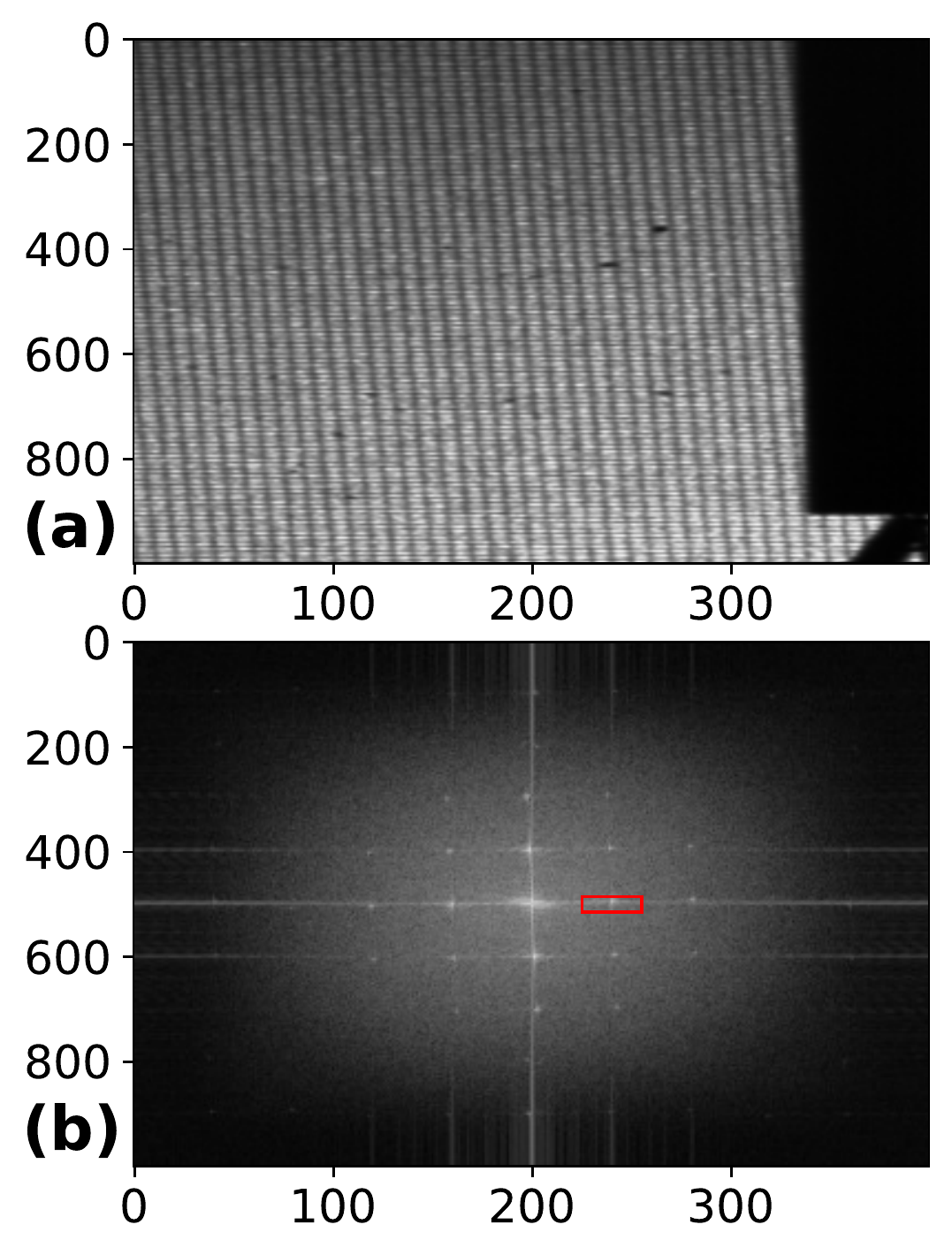}
    \caption{(a) Experimentally recorded image which serves as reference image. (b) Fourier spectrum. Marked area shows region of interest }
    \label{fig:ref}
\end{figure}

Many optical techniques are based on grids, such as moiré \cite{sciammarella1982moire,amidror2009theory,wang2022sampling} and fringe projection \cite{gorthi2010fringe}. Grids are usually explicit, that is visually evident. However, grids can also be implicit, such as the structures of CCD detectors and LCD modulators. Hidden grids, proposed in moiré metrology \cite{garavaglia2001optoelectronic}, are applied in BOS in this paper.
A schematic setup of a hidden grid background oriented schlieren is shown in Figure \ref{fig:1}.

The light source is a video projector, which is used to project a white image onto a ground glass, acting as background screen. Because of the pixel structure of the projector, a hidden grid is present on the ground glass. The test section consisted of a central (heated) plate and two shrouding (unheated) vertical walls. 
The central plate was heated by electrical plane resistances and instrumented with chromel-alumel thermocouples, calibrated to $\pm$ 0.1 K, more details are given in \cite{ambrosini2005comparative}.
The phenomenon investigated is natural convection in air. 
A CMOS TV camera (Silicon Video 9T001C, resolution 2048 x 1536 pixels, 3.2 $\mu$m x 3.2 $\mu$m pixel size) captures the background grid. The camera is equipped with a TEC-55 55 mm F/2.8 Telecentric Computar Lens.

The mathematical form of the recorded grid pattern can be expressed as \cite{grediac2016grid}
\begin{align}
    I(x,y) &= a(x,y)+b(x,y)\mathcal{F}(\omega_{cx}x+\phi(x,y)) \notag\\
    &+ c(x,y)\mathcal{F}(\omega_{cy}y+\psi(x,y))
\end{align}
where $a(x,y)$ is the background intensity, $b(x,y)$ and $c(x,y)$ denote fringe amplitudes, $\mathcal{F}$ is a real periodic function, and $(\omega_{cx},\omega_{cy})$ denote the angular spatial carrier frequencies along horizontal and vertical dimensions.
In the above equation, the periodic function $\mathcal{F}$ can be visualized as the sum of several sinusoids using the Fourier series decomposition.
Further, $\phi$ and $\psi$ are the phase modulation terms along the two dimensions.
A sample experimentally recorded hidden grid image is shown in Figure \ref{fig:ref}(a).
The corresponding normalized Fourier spectrum (logarithmic values) is shown in Figure \ref{fig:ref}(b).
It is evident that we have distinct multiple lobes or regions in the spectrum.
For our data processing, we applied geometric phase analysis method \cite{dai2014geometric} where we select a single lobe or region of interest (marked in red color) in the Fourier spectrum plot of Figure \ref{fig:ref}(b) for phase extraction.
This is achieved by using the local fringe processing property of the windowed Fourier transform. Accordingly, we compute the windowed Fourier transform of the signal $I(x,y)$ as \cite{kemao2007two}
\begin{align}
    S(u,v;\xi,\eta) = \int_{-\infty}^{\infty} \int_{-\infty}^{\infty} &I(x,y) w(x-u,y-v)\\\notag
    & e^{-i(\xi x + \eta y)}\,dx dy
\end{align}
where $(u,v)$ represent spatial coordinates and $(\xi,\eta)$ denote spatial frequency coordinates.
In other words, the quantity $S$ is a function of both space and frequency, and is the result of a space-frequency transform.
Further, $w(x,y)$ is a real symmetric window function and is usually chosen to be a Gaussian function.
Mathematically, the normalized Gaussian window function can be expressed as
\begin{equation}
    w(x,y) =  \exp\left(-\frac{x^2}{2\sigma_x^2}-\frac{y^2}{2\sigma_y^2} \right)
\end{equation}
The size of the window is characterized by the parameters $\sigma_x$ and $\sigma_y$.
The local spatial frequencies $\omega_x$ and $\omega_y$ along the horizontal and vertical dimensions are computed as,
\begin{equation}
    [\omega_x(u,v), \omega_y(u,v)] = \arg\max_{\xi,\eta} |S(u,v;\xi,\eta)|
\end{equation}
In the above equation, $[\omega_x(u,v), \omega_y(u,v)]$ indicates location of the peak of the windowed Fourier spectrum.
The peak is also referred to as windowed Fourier ridge and the phase is computed as,
\begin{multline} \label{eq: phase}
    \phi_0(u,v) = \text{angle}\{S_0[u,v;\omega_x(u,v),\omega_y(u,v)]\} \\+ \omega_x(u,v)u + \omega_y(u,v)v
\end{multline}
The main advantage of the windowed Fourier transform method is the feasibility of jointly studying the signal characteristics with respect to both the spatial and frequency domains.
Note that the WFT method also provides a filtering algorithm, usually referred to as windowed Fourier filtering (WFF) algorithm, for denoising the fringe pattern using a preset threshold parameter value \cite{kemao2007two}.
However, as we were mainly interested in phase extraction and not fringe filtering, we used the windowed Fourier ridge (WFR) algorithm for our analysis.
For heat flow investigations, we primarily considered the dimension transverse to the heated plate, or the horizontal dimension, and hence were mainly interested in the phase modulation term $\phi(x,y)$ in our computations.
In the windowed Fourier transform method, the local spectral processing aspect is controlled by the parameters $(wxl,wxh)$ and $(wyl,wyh)$ which indicate the range of angular spatial frequencies along the horizontal and vertical dimensions \cite{kemao2005two}.
The step sizes for the two ranges of frequencies are denoted by $wxi$ and $wyi$.
In our analysis, we used $wxl=0.5,wxi=0.01,wxh=1$ and $wyl=0.0,wyi=0.00025,wyh=0.001$ for the windowed Fourier ridge algorithm.
Note that the layout of the hidden grid, including the grid spacing, affects the image spectrum and location of spatial frequencies. 
Hence, the choice of the WFT parameters such as local frequency range would be affected by the hidden grid features, which depend on the experimental configuration. 
In addition, the window size parameter in WFT method is mainly decided by noise in the fringe signal since a small window offers better local phase approximation used in the WFT algorithm whereas a large window is more robust against noise \cite{kemao2007two}. 
However, a window size parameter of $\sigma_x=\sigma_y=10$ pixels has been suggested for most practical cases \cite{kemao2007two}, and we used the same window size parameter in our analysis.

\section{Results}
\begin{figure*}[t] 
    \centering
    {\includegraphics[width=0.95\textwidth]{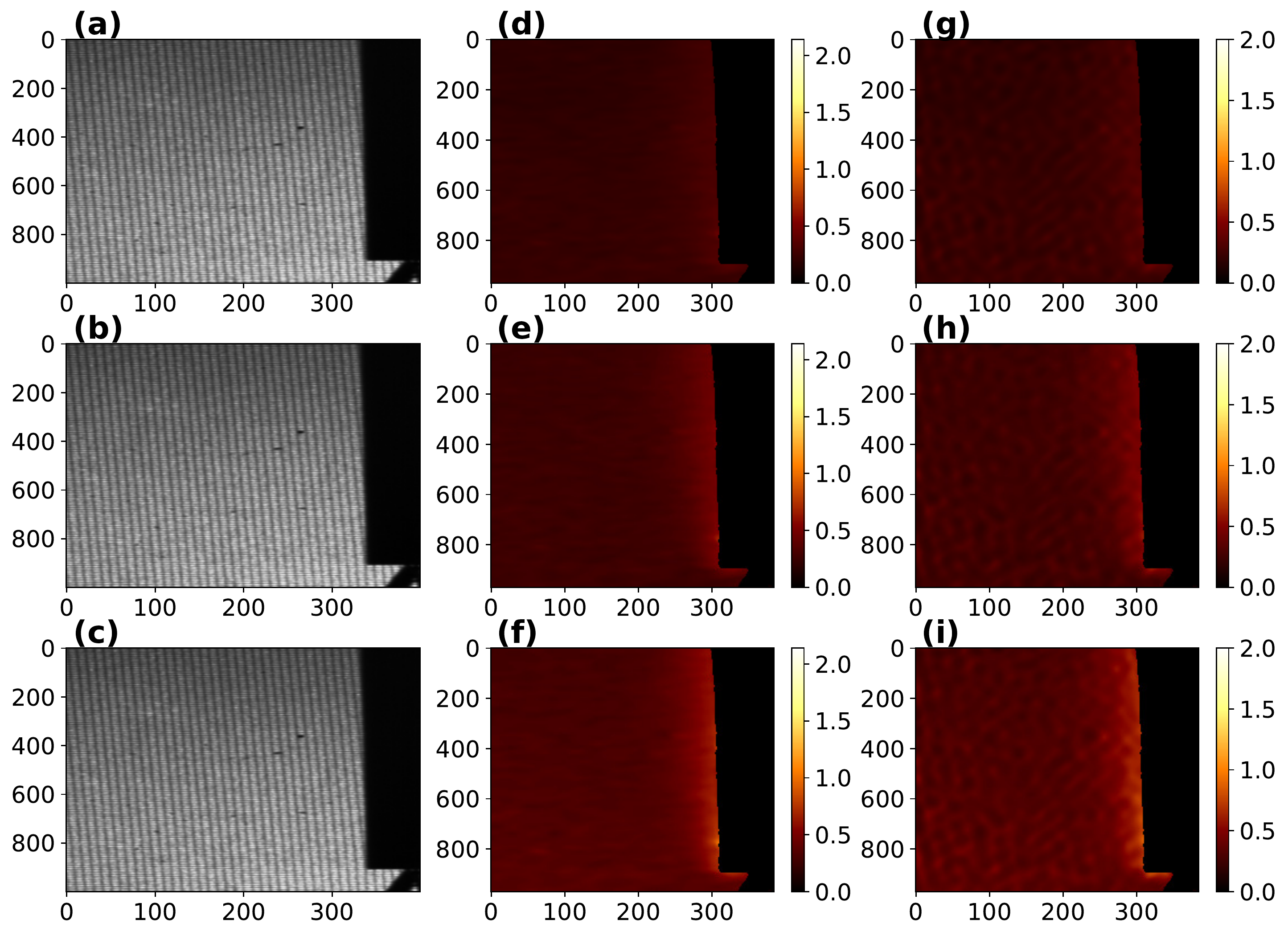}}
    \caption{
        Data processing for Frames 1 to 3. 
        (a-c) Experimentally recorded intensity patterns.
        (d-f) Phase maps estimated using windowed Fourier transform method.
        (g-i) Phase maps estimated using Fourier transform method.
        The phase values are in radians and horizontal and vertical axes labels indicate the pixel coordinates.
    }
    \label{fig:frame1}
\end{figure*}

\begin{figure*}[h!]
    \centering
    {\includegraphics[width=0.95\textwidth]{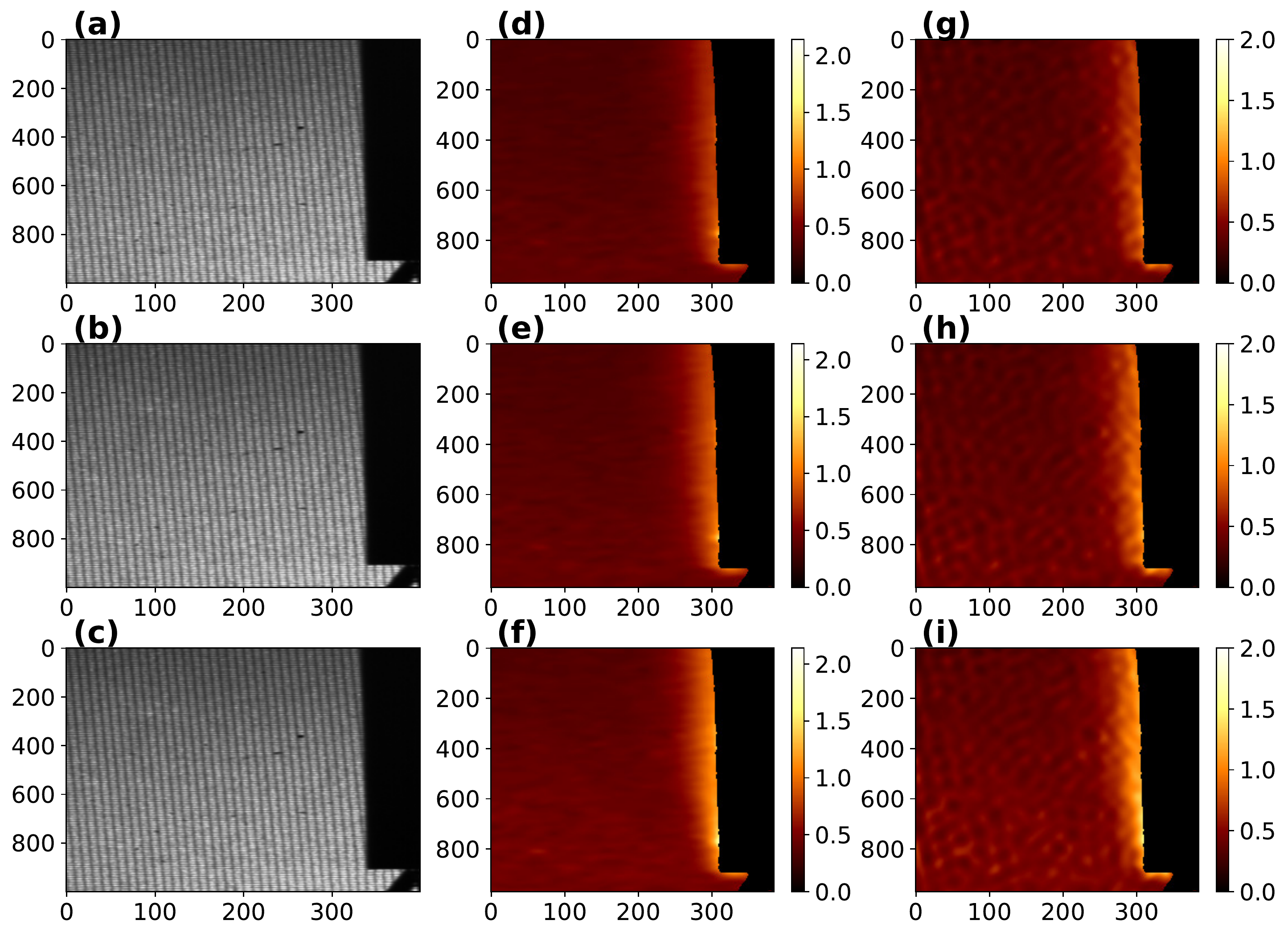}}
    \caption{
        Data processing for Frames 4 to 6. 
        (a-c) Experimentally recorded intensity patterns.
        (d-f) Phase maps estimated using windowed Fourier transform method.
        (g-i) Phase maps estimated using Fourier transform method.
        The phase values are in radians and horizontal and vertical axes labels indicate the pixel coordinates.
    }
    \label{fig:frame2}
\end{figure*}

\begin{figure*}[t!]
    \centering
    {\includegraphics[width=0.95\textwidth]{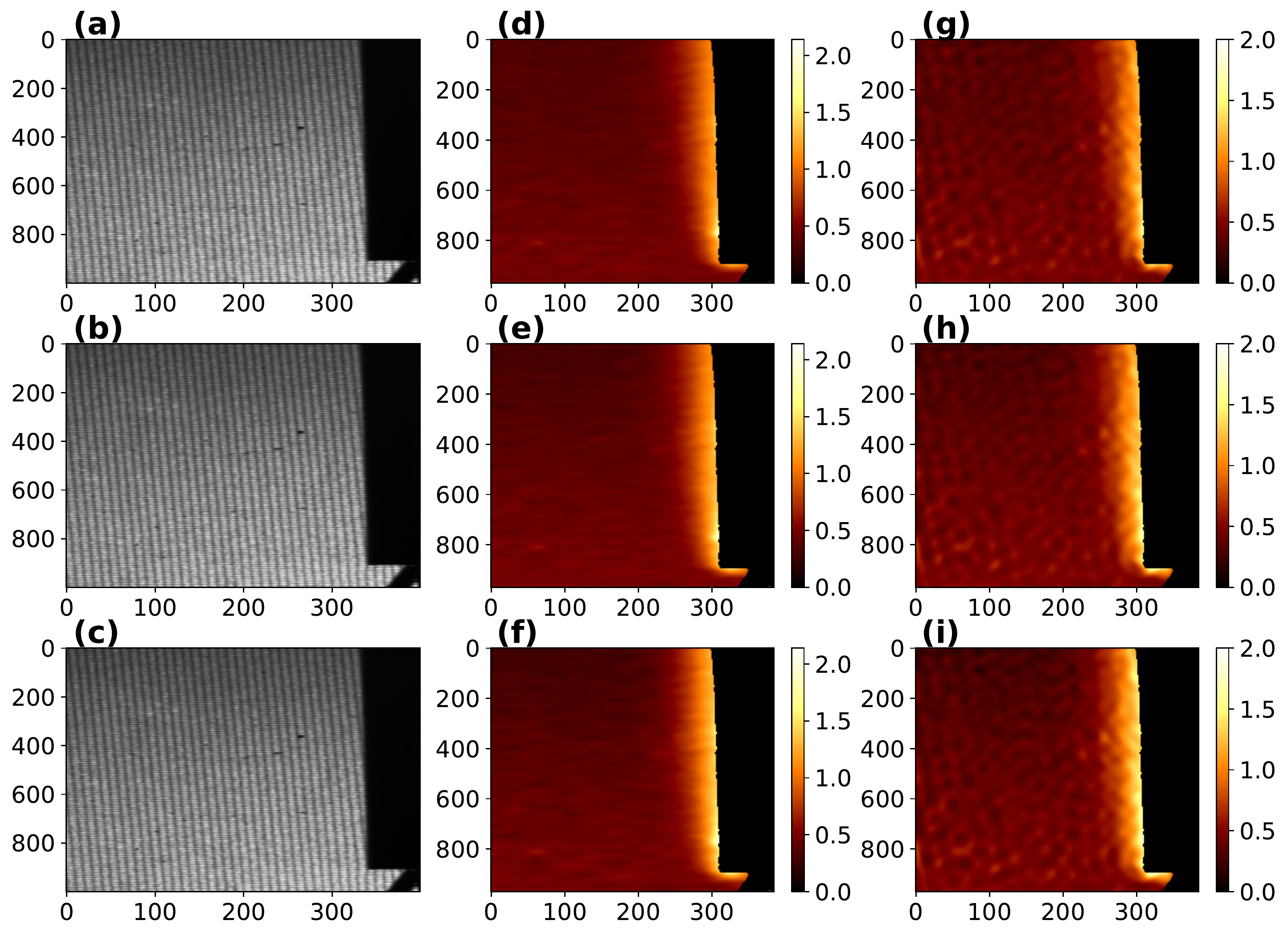}}
    \caption{
        Data processing for Frames 7 to 9. 
        (a-c) Experimentally recorded intensity patterns.
        (d-f) Phase maps estimated using windowed Fourier transform method.
        (g-i) Phase maps estimated using Fourier transform method.
        The phase values are in radians and horizontal and vertical axes labels indicate the pixel coordinates.
    }
    \label{fig:frame3}
\end{figure*}

\begin{figure*}[h!]
    \centering
    {\includegraphics[width=\textwidth]{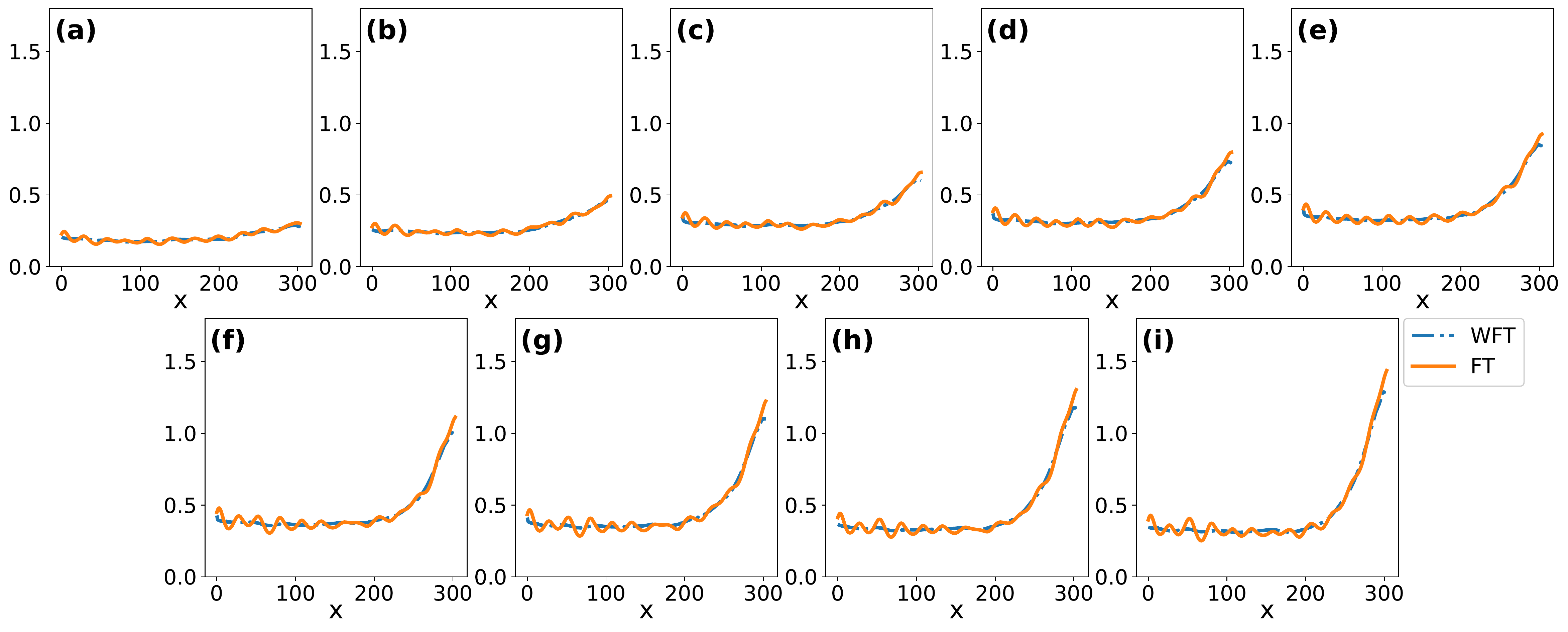}}
    \caption{(a-i) Line profile plots along row number 200 for the estimated phase maps in radians from Frame 1 to 9. }
    \label{fig:line1}
\end{figure*}

\begin{figure*}[t!]
    \centering
    {\includegraphics[width=\textwidth]{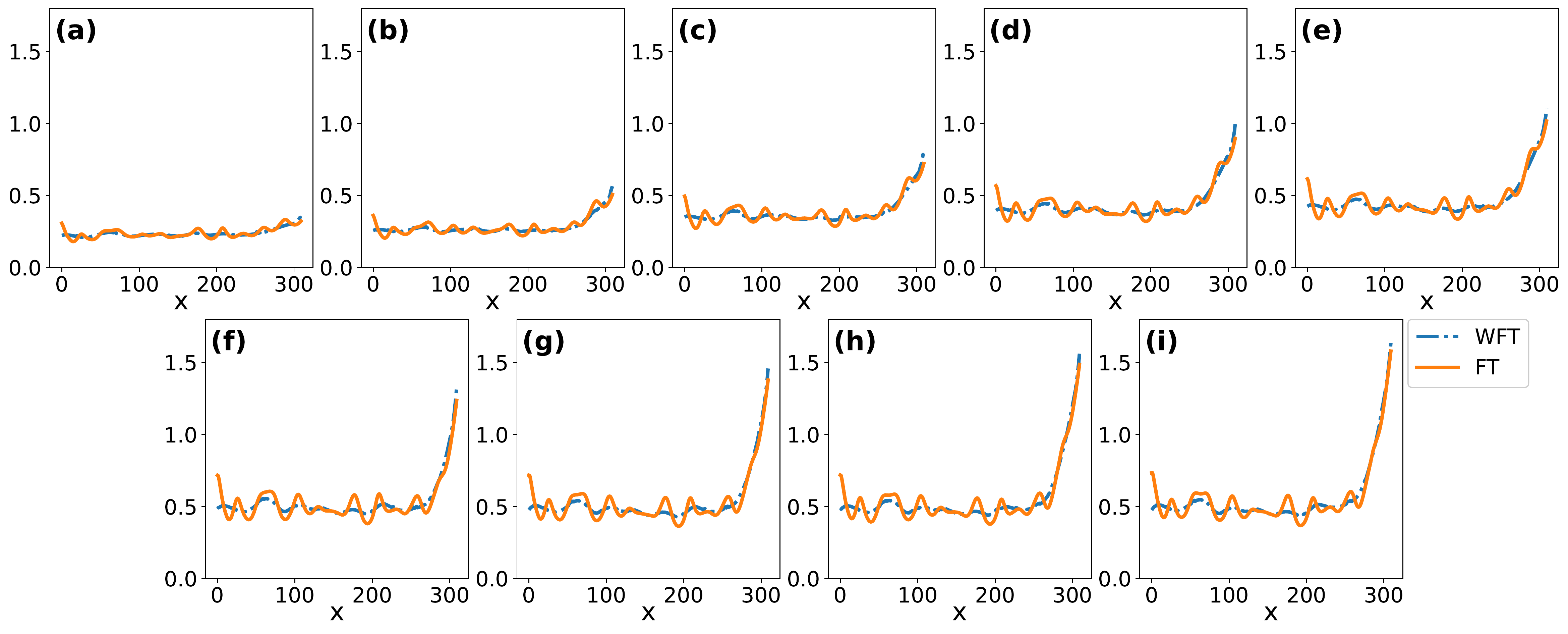}}
    \caption{(a-i) Line profile plots along row number 800 for the estimated phase maps in radians from Frame 1 to 9. }
    \label{fig:line4}
\end{figure*}
\begin{figure}[h!]
    \centering
    \includegraphics[width=0.37\textwidth]{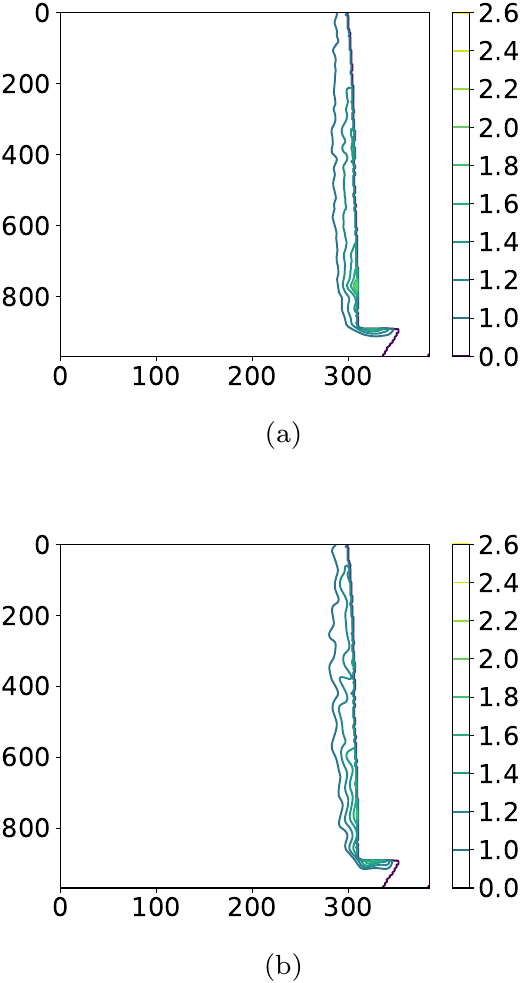}
    \caption{Phase contour plot in radians obtained using (a) windowed Fourier transform method and (b) Fourier transform method.}
    \label{fig:contour}
\end{figure}
\begin{figure}[h!] 
    \centering
    {\includegraphics[width=0.4\textwidth]{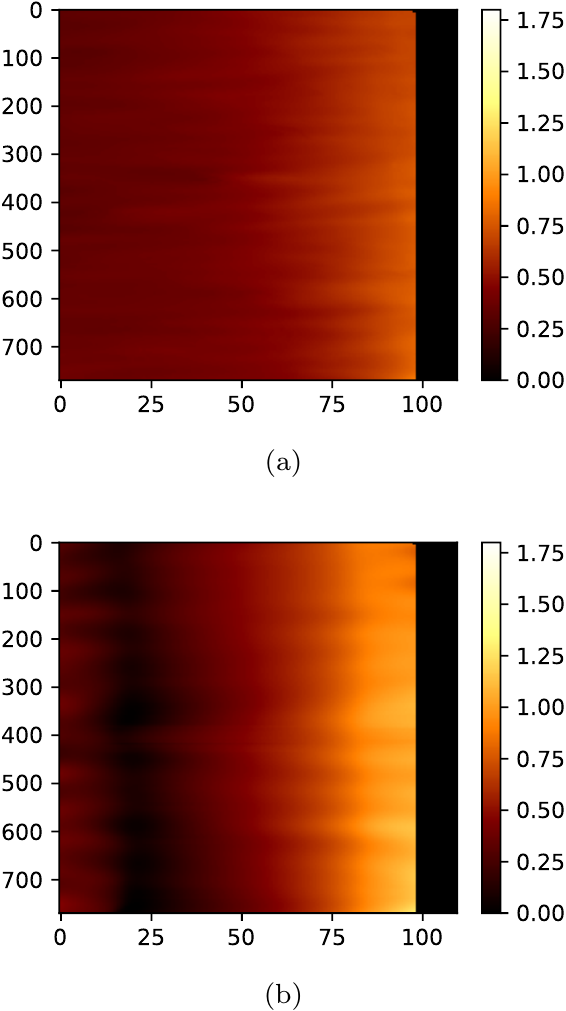}}
    \caption{
        Phase maps in radians for region near the heated plate structure using (a) windowed Fourier transform method and (b) wavelet transform method.
    }
    \label{fig:wavelet}
\end{figure}

In our experiment, we studied convective heat transfer in vertical channels. The test section \cite{ambrosini2005comparative} was made by a central (heated) Aluminium plate, shrouded by two unheated vertical walls. 
The heated plate had the following dimensions: overall thickness $0.012$ m, height $0.175$ m, length $0.3$ m. The shrouding walls had the same height and length as the heated plate.
An image corresponding to the initial state, that is with no heating applied to the plate, was first captured. Then we heated the central plate and the process of heat transfer by convection began. The change in refractive index, caused by the temperature gradient, induced a phase modulation in the grid pattern, seen through the test cell. 
Different images were recorded at different temperatures of the heated plate. 
For our analysis, ten intensity patterns (Frame 0 to 9) were processed in the experiment.
The first frame (Frame 0) is considered as the reference frame against which the phase variations were measured. It was recorded with the plate in thermal equilibrium with ambient air at temperature $T=293$ K.
Subsequent frames were recorded increasing the heated plate temperature with steps of $3$ K (i.e. frame 1 at $T=296$ K, frame 2 at $T=299$ K etc.).

The reference frame is shown in Figure \ref{fig:ref}(a).
The reference frame approach also removes the effect of spatial carrier and background artifacts \cite{colomb2006total}.
Using the windowed Fourier transform method, we estimated the phase maps from each frame and computed the difference phase with respect to the reference phase.
The dark region corresponding to the heated plate was ignored in our analysis by applying a binary mask.
In addition, we also neglected the pixels close to the masked region to ignore the effect of boundary errors.

The intensity patterns corresponding to Frames 1 to 3 are shown in Figures \ref{fig:frame1}(a-c).
Note that in these images, the x-axis indicates the horizontal dimension and the y-axis indicates the vertical dimension.
The estimated phase maps for these frames using the windowed Fourier transform method are shown in Figures \ref{fig:frame1}(d-f).
For comparison, we also computed the phase maps using the standard Fourier transform method, whose results are shown in Figures \ref{fig:frame1}(g-i).
Similar results for Frames 4 to 6 are shown in Figure \ref{fig:frame2} and for Frames 7 to 9 are shown in Figure \ref{fig:frame3}.

From the figures, we can observe that each phase map exhibits spatial variations with respect to the distance from the heated plate.
We observe regions of high phase values near the plate and there is a gradual decrease in phase as we move away from the heated structure.
Hence, by measuring the phase map, our approach enables quantitative visualization of the heat flow.
Further, we also observe that the peak phase value changes with the frame number indicating the dynamic nature of heat transfer. 
The connection between phase values and temperature gradient is described in \cite{ambrosini2007heat,rajshekhar2018multi}.

The comparative results also show that the windowed Fourier transform provides a smoother map in comparison with the Fourier transform method.
For better quantitative assessment, we also computed the line profiles of the phase maps corresponding to two rows of the phase map for each frame.
For row number 200, the line profile plots for different frames corresponding to the windowed Fourier transform method and the Fourier transform method are shown in Figure \ref{fig:line1}.
Similarly, for row number 800, the line profile plots for different frames corresponding to the two methods are shown in Figure \ref{fig:line4}.
The line profiles also indicate that the phase obtained from Fourier transform method exhibits more fluctuations and is relatively non-smooth  as compared to the phase obtained from the windowed Fourier transform method. This is particularly evident in Figure \ref{fig:line4}, as line 800 is near to the edge of the image.

In addition, for ease of visualization, we also show the contour plots of phase maps corresponding to Frame 8  in Figure \ref{fig:contour}.
The contour plot for phase in radians obtained using the windowed Fourier transform method is shown in Figure \ref{fig:contour}(a).
Similarly, the contour plot for phase obtained using the Fourier transform method is shown in Figure \ref{fig:contour}(b).
From these figures, we observe that windowed Fourier transform method outperforms the Fourier transform method.
Further, the WFT results are also in good agreement with the results obtained by focal filament method and standard schlieren on the same test section \cite{ambrosini2005comparative}.

We also show comparison with wavelet transform method \cite{watkins1999determination} for phase extraction corresponding to Frame 4 in Figure \ref{fig:wavelet}.
In the figure, we used a small rectangular mask for the heated plate, and show the phase maps in a region near the heated plate.
The phase map obtained using the windowed Fourier transform method is shown in Figure \ref{fig:wavelet}(a).
Similarly, the phase map obtained using the wavelet transform method is shown in Figure \ref{fig:wavelet}(b).
We observe that windowed Fourier transform offers smoother phase map with relatively less artifacts as compared to the wavelet transform method.

\section{Discussion}

The results shows that hidden grid background oriented schlieren has the same capabilities of standard BOS in studying flow field. In particular, the technique clearly identifies the thermal boundary layer, that is the region of flow, near the surface, in which the temperature gradients (and hence the phase variations) are significant. 
The proposed setup exhibits less noise, with respect to BOS systems using a laser source. Furthermore, illumination is very efficient (a white image is projected), therefore high f-numbers can be used yielding high depth of field, useful to keep in focus both the background and the test section.
Our results also show that windowed Fourier transform method has strong potential as robust data processing technique for visualizing heat flows.
The localized processing capability offered by the method leads to better noise robustness and less susceptibility against local image abnormalities. 
The main limitation of the windowed Fourier transform method is the high computational cost due to joint processing of space and frequency components.
However, in recent years, several efforts have been documented to improve the computational efficiency of the method using parallel processing techniques \cite{gao2012parallel,wang2018parallel}.
These techniques offer exciting opportunities for massive reduction in computational overhead associated with the windowed Fourier transform method.

\section{Conclusion}

In the article, we introduced hidden grid background oriented schlieren and demonstrated its practical application, coupled to data processing by windowed Fourier transform, for quantitative heat flow study.
The authors believe that the method offers great scope to enhance the experimental utility of these techniques and could offer interesting insights about the dynamics of heat transfer.  

\section{Acknowledgment}
Gannavarpu Rajshekhar gratefully acknowledges the funding obtained from Department of Science and Technology(DST), India under grant number DST/NM/NT/2018/2.


\end{document}